\documentclass[12pt]{article}
\usepackage{graphicx}

\begin{document}

\author{W-Y. P. Hwang and Chih-Yi Wen \\
%EndAName
The Leung Research Center for Cosmology and Particle Astrophysics,\\
Center for Theoretical Sciences,\\
Department of Physics, and Institute of Astrophysics,\\
National Taiwan University,\\
Taipei, Taiwan 106, ROC}
\title{Parity-Violating Nuclear Force \\
as derived from QCD Sum Rules}
\date{January 26, 2008}
\maketitle

\begin{abstract}
Parity-violating nuclear force, as may be accessed from parity
violation studies in nuclear systems, represents an area of
nonleptonic weak interactions which has been the subject of
experimental investigations for several decades. In the simple
meson-exchange picture, parity-violating nuclear force may be
parameterized as arising from exchange of $\pi$, $\rho$, $\omega$,
or other meson(s) with strong meson-nucleon coupling at one vertex
and weak parity-violating meson-nucleon coupling at the other
vertex. The QCD sum rule method allows for a fairly complicated,
but nevertheless straightforward, leading-order loop-contribution
determination of the various parity-violating $MNN$ couplings
starting from QCD (with the nontrivial vacuum) and
Glashow-Salam-Weinberg electroweak theory. We continue our earlier
investigation of parity-violating $\pi NN$ coupling (by Henley,
Hwang, and Kisslinger) to other parity-violating couplings. Our
predictions are in reasonable overall agreement with the results
estimated on phenomenological grounds, such as in the now classic
paper of Desplanques, Donoghue, and Holstein (DDH), in the global
experimental fit of Adelberger and Haxton (AH), or the effective
field theory (EFT) thinking of Ramsey-Musolf and Page (RP).

\vskip 1.5 true cm
\parindent=0truept
PACS: 24.80.+y (Nuclear tests of fundamental interactions and
symmetries), 24.85.+p (Quarks, gluons, and QCD in nuclei and
nuclear processes).

\end{abstract}

\section{Introduction}

Parity violation (PV) studies in nuclear systems, such as the
asymmetry\cite{vanOers,Carlson} in ${\vec p}+p \to p+p$, the
photon circular polarization measurement in $n+p\to d+\gamma $,
and PV studies in $^{18}F$ and $^{21}Ne$, offer a means of
determining the parity-violating nuclear force, which represents a
special category of nonleptonic weak interactions accessible
experimentally. In the simple meson-exchange picture,
parity-violating nuclear force arises from exchange of $\pi $,
$\rho $, $\omega $, or other meson(s) with strong $MNN$ coupling
at one vertex and weak parity-violating $MNN$ coupling at the
other vertex. In 1980's, parity-violating meson-nucleon couplings
have been estimated primarily on phenomenological grounds, such as
in the classic paper of Desplanques, Donoghue, and Holstein (DDH)
\cite{DDH}. A global fit to obtain these PV meson-nucleon
couplings, making use of the various experimental data available
at the time, was performed also (in 1985) by Adelberger and Haxton
(AH) \cite{AH}. Nevertheless, progress\cite{Michael}, both
experimental and theoretical, has been slow since then, although
the interest in the problem arose to some extent in view of the
adopted effective field theory for the few nucleon systems.

In 1980's, the standard model of particle physics, which consists
of Glashow-Salam-Weinberg [GSW] electroweak theory and quantum
chromodynamics [QCD], had meanwhile been well established.
Although the nonperturbative feature of QCD manifests itself in
formation of hadron structure and makes it very difficult to
predict quantitatively hadron properties including both the strong
parity-conserving and weak parity-violating meson-nucleon
couplings, the method of QCD sum rules\cite {SVZ} offers a
systematic way for taking into account effects as caused by the
nontrivial nature of the QCD ground state or the QCD vacuum. The
QCD sum rule method has become quite popular starting from 1990's
(till now). In view of the laborious efforts in lattice gauge
theory, the QCD sum rule method does offer an alternative avenue
to obtain the various predictions, to some extent in the spirit of
QCD.

To "complete" the study of nuclear parity violations, we need to
investigate both the parity-conserving and parity-violating
meson-nucleon couplings, using the conventional scheme of nuclear
forces. As already shown in another paper \cite{wen96a}, it is
possible to use the method of QCD sum rules in external fields
\cite{Ioffe} to determine the strong $\pi NN$, $\rho NN$, and
$\omega NN$ couplings. The purpose of this paper is to present a
QCD sum rule determination of the parity-violating $\rho NN$ and
$\omega NN$ couplings, which turns out to be fairly complicated
but nevertheless straightforward. Our present results, together
with a previous study of the parity-violating $\pi NN$ coupling
\cite{HHK}, allow for a direct connection between parity-violating
nuclear force and the standard model, despite the complication of
QCD.

The external-field QCD sum rule method\cite{Ioffe} have been used
to treat the strong $\pi NN$ coupling \cite{Hwang0}, the weak
parity-violating $\pi NN$ coupling \cite{HHK}, and the strong
$\rho NN$ and $\omega NN$ couplings \cite{wen96a}. In all cases
considered, quantitative successes have been achieved mainly
because the nonperturbative effects of QCD, as expressed in terms
of induced condensates, have been taken into account and are found
to be of critical importance. Our efforts to treat hyperon weak
decays remain in progress \cite{Henley}, with fairly encouraging
results (which are beyond the scope of the present article).
Therefore, it is natural to follow the same method to treat the
weak parity-violating $\rho NN$ and $\omega NN$ couplings.
Nevertheless, it remains difficult to assess properly whether our
present calculation may also succeed in a quantitative manner,
since after all the nonleptonic amplitudes in question are
``notoriously difficult to calculate'' (quoting the phrase from
the referee on the early version of one of my early articles).
However, we have good reasons, based on successful experiences on
the method of QCD sum rules in general, to believe that the
complicated task carried out in this research could be the first
important step in establishing a benchmark in the (future)
quantitative treatments of this difficult problem. Just like many
other nonleptonic weak decays where relevant data exist, we
already have experimental information \cite{AH,Michael} which we
may use to test our theoretical predictions.

For the sake of completeness, we begin by outlining a few
ingredients regarding the external-field QCD sum rule method,
without detailed qualifying statements. The external vector field is
expressed as
\begin{equation}
Z_{\mu }=-\frac{1}{2}Z_{\mu \nu }x^{\nu }.
\end{equation}
We attempt to determine the following polarization function for a
nucleon in a small (classical) external vector-meson field $Z_{\mu
}$ (where $Z$ represents either a $\rho $ or $\omega $ meson):
\begin{equation}
i\int d^{4}x\ e^{iq\cdot x}\left\langle 0|T\eta _{N}(x)\bar{\eta}%
_{N}(0)|0\right\rangle _{Z}=\Pi (q)+hZ_{\mu \nu }\Pi ^{\mu \nu }(q),
\end{equation}
where the numerical constant $h$ is the coupling between the
external field and the up ($u$) quark field; more explicitly, the
coupling between the external field and the down ($d$) quark is $-h$
for $\rho $ meson and $+h$ for $\omega $ meson. The standard form
for the composite operator $\eta _{N}(x)$ is adopted \cite{Ioffe2}:

\begin{equation}
\eta _{p}(x)=\epsilon ^{abc}[u^{aT}(x)C\gamma _{\mu
}u^{b}(x)]\gamma _{5}\gamma^{\mu }d^{c}(x),
\end{equation}

\begin{equation}
\eta _{n}(x)=\epsilon ^{abc}[d^{aT}(x)C\gamma _{\mu
}d^{b}(x)]\gamma _{5}\gamma^{\mu }u^{c}(x),
\end{equation}
which transform like the proton and neutron fields, respectively.
Here $u^{a}(x)$ and $d^{a}(x)$ are the up and down quark fields with
the superscript $a$ the color index and $C$ is the charge
conjugation operator.

At the hadronic level, we define the parity-violating
meson-nucleon couplings related to $\rho$ and $\omega$ mesons in
the standard manner\cite{DDH}.

\begin{eqnarray}
L_{int}^{p.v.} &=&-\bar{N}[h_{\rho }^{0}\vec{\tau}\cdot
\vec{\phi}_{\mu }^{\rho }+h_{\rho }^{1}\phi _{\mu }^{\rho
3}+\frac{h_{\rho }^{2}}{2\sqrt{6}} (3\tau ^{3}\phi _{\mu }^{\rho
3}-\vec{\tau}\cdot \vec{\phi}_{\mu }^{\rho })
]\gamma ^{\mu }\gamma _{5}N  \nonumber \\
&&+h_{\rho }^{^{\prime }1}\bar{N}(\vec{\tau}\times \vec{\phi}_{\mu
}^{\rho })^{3}\frac{\sigma ^{\mu \nu }k_{\nu }}{2m}\gamma
_{5}N-\overline{N}[ h_{\omega }^{0}\phi _{\mu }^{\omega }+h_{\omega
}^{1}\tau ^{3}\phi _{\mu }^{\omega }]\gamma ^{\mu }\gamma _{5}N.
\end{eqnarray}
Here the superscripts in the couplings $h_\rho^0$, $h_\rho^1$,
$h_\rho^2$, $ h_\rho^{\prime 1}$, $h_\omega^0$, and $h_\omega^1$
refer to the isospin character as the weak interactions do not
observe isospin symmetry. We may use the above expression to write
down the polarization function $\Pi(q)$, with the paricular piece
which is proportional to $h$, at the hadronic level. It is clear
that we need to combine different channels (i.e. $p\to p$, $n\to n$,
$p\to n$, and $n\to p$) in order to extract these PV coupling
constants.

At the quark level, we use weak interactions as described by the
Glashow-Weinberg-Salam (GSW) electroweak theory to determine the
polarization function, obtaining in general three-loop diagrams
which require regularizations. Here we adopt dimensional
regularization in the minimum-subtraction (MS) scheme and introduce
suitable counter terms in defining the renormalized operators $[\eta
_{p}]_{R}$ and $[\eta _{n}]_{R}$. As mentioned above,  we may
parametrize the same polarization function phenomenologically,
making use of the parity-violating $\rho NN$ and $\omega NN$
interaction Lagrangians. Comparing the results obtained through the
two ways of evaluating the polarization functions (i.e., at the
quark level using GSW theory and QCD and at the hadronic level
involving $\rho NN$ and $\omega NN$ couplings), we have a definitive
way in extracting the weak PV $\rho NN$ and $\omega NN$ couplings
$-$ the primary objective of this paper.

\section{The parity-violating $\rho NN$ and $\omega NN$ couplings}

As a special feature in relation to the method of QCD sum rules, the
kinematic variable $q_\mu^2$ of the polarization function $\Pi(q)$,
as translated into the choice of the Borel mass squared $M^2$, which
is in the vicinity of slightly above $1\, GeV^2$. Such choice of the
Borel mass is to ensure the approximate validity of the
operator-product expansion (OPE) augmented with power corrections
(as due to the various condensates). In other words, perturbative
QCD corrections to the coefficients in such OPE are in principle
there but are presumably suppressed by choice of the Borel mass $M$.
Unlike what has been involved in most phenomenological approaches to
the problem where some effective weak Hamiltonian at the energy
scale relevant to the hadron must be directly invoked, we have in
the QCD sum rule method the nice feature that the GSW electroweak
theory is called for at the scale set by the Borel mass squared
$M^2$ where effects to order $O(G_F\alpha_S)$ are suppressed (due to
the running of the strong coupling $\alpha_S$) and it is the
intrinsic smooth extrapolation of the results to lower $q^2$ which
helps to explain the successes of the predictions.

Accordingly, it should be possible to use, in the QCD sum rule
method, the effective GSW electroweak lagrangian at tree level while
leaving terms in $O(G_F\alpha_S)$ as corrections. That is, we may
use, as a good starting point,

\begin{eqnarray}
L_{weak} &=&-\frac{G_{F}}{\sqrt{2}}[\bar{u}\gamma _{\mu }(1-\gamma _{5})d%
\bar{d}\gamma^{\mu }(1-\gamma _{5})u+  \nonumber \\
&&\sum_{q_{1},q_{2}=u,d}\bar{q}_{1}\gamma _{\mu
}(A_{q_{1}}-B_{q_{1}}\gamma _{5})q_{1}\bar{q}_{2}\gamma^{\mu
}(A_{q_{2}}-B_{q_{2}}\gamma _{5})q_{2}],
\end{eqnarray}
where $G_F$ is the Fermi coupling constant ($G_{F}=1.166\times
10^{-5}\,GeV^{-2}$) and the other constants are defined through
$A^{u}=\frac{1}{2}-\frac{4}{3}\sin^{2}\theta _{W}$ ,
$B^{u}=\frac{1}{2}$ , $A^{d}=-\frac{1}{2}+\frac{2}{3}\sin^{2}\theta
_{W}$ and $B^{d}=-\frac{1}{2}$ with $sin^2\theta _{W}$ the
electroweak mixing parameter ($sin^2\theta_W= 0.2315$). Note that
the first term in the above equation comes from exchange of the
$W^{\pm }$ boson while the second term is due to $Z^{0}$ exchange.
In view of our primary goal which is to identify the sizable role
played by the various condensates, it is clear that the issue
regarding the renormalization of the tree lagrangian (as would be
suppressed by the choice of the Borel mass) is of secondary
importance and could be addressed when higher precision (in
theoretical prediction) is called for. As shall be explained later,
it is in fact sufficient to even use the current-current form since
the difference from what we may obtain by employing the
renormalizable gauge throughout, such as the $R_\xi$ gauge, does not
matter at the end as far as our final QCD sum rules are concerned.

\subsection{Renormalization of composite operators}

Inclusion of the weak interaction in the polarization function leads
to sub-divergences even in the lowest order $O(G_{F})$. Such
divergences may be removed (or regularized) by making use of a
suitably defined renormalized composite operators $[\eta ]_{R}$ and
$[\bar{\eta}]_{R}$. Such renormalized operators may be obtained by
considering the four-fermion interaction:

\begin{eqnarray}
&&\int d^{d}x\ d^{d}y\ e^{i(p\cdot x+q\cdot y)}\left\langle
u_{i^{^{\prime
}}}^{a^{^{\prime }}}(0)d_{j^{^{\prime }}}^{b^{^{\prime }}}(0)\bar{u}%
_{i}^{a}(x)\bar{d}_{j}^{b}(y)\right\rangle =\frac{\delta
^{a^{^{\prime
}}a}\delta ^{b^{^{\prime }}b}}{p^{2}q^{2}}\hat{p}_{i^{^{\prime }}i}\hat{q}%
_{j^{^{\prime }}j}+  \nonumber \\
&&\qquad \frac{G_{F}}{\sqrt{2}}\frac{1}{48\pi ^{2}(d-4)}\frac{1}{p^{2}q^{2}}[%
(p+q)_{\rho }(p+q)_{\sigma }+\frac{1}{2}(p+q)^{2}g_{\rho \sigma
}]\times
\nonumber \\
&&\qquad \{[\gamma ^{\rho }\gamma _{\mu }(A^{u}-B^{u}\gamma _{5})\hat{p}%
]_{i^{^{\prime }}i}[\gamma ^{\sigma }\gamma ^{\mu }(A^{d}-B^{d}\gamma _{5})%
\hat{q}]_{j^{^{\prime }}j}+  \nonumber \\
&&\qquad [\gamma ^{\rho }\gamma _{\mu }(1-\gamma
_{5})\hat{p}]_{i^{^{\prime }}i}[\gamma ^{\sigma }\gamma ^{\mu
}(1-\gamma _{5})\hat{q}]_{j^{^{\prime }}j}\}+finite\ parts,
\end{eqnarray}
where the first term is the free piece while the second term, which
is divergent at $d=4$, comes from the insertion of the weak
interaction. This equation enables us to define the renormalized
operator $[u(x)d(x)]_{R}$. Using the minimum-subtraction (MS)
scheme, we obtain  the renormalized operator $[u(x)d(x)]_{R}$ as
follows:

\begin{eqnarray}
\left[ u_{i}^{a}(x)d_{j}^{b}(x)\right] _{R} &=&u_{i}^{a}(x)d_{j}^{b}(x)+%
\frac{G_{F}}{\sqrt{2}}\frac{\mu ^{d-4}}{48\pi ^{2}(d-4)}(\partial
_{\rho
}\partial _{\sigma }+\frac{1}{2}g_{\rho \sigma }\bar\sqcup )\times  \nonumber \\
&&\{[\gamma ^{\rho }\gamma _{\mu }(A^{u}-B^{u}\gamma
_{5})u^{a}(x)]_{i}[\gamma ^{\sigma }\gamma ^{\mu }(A^{d}-B^{d}\gamma
_{5})d^{b}(x)]_{j}+  \nonumber \\
&&[\gamma ^{\rho }\gamma _{\mu }(1-\gamma _{5})u^{a}(x)]_{i}[\gamma
^{\sigma }\gamma ^{\mu }(1-\gamma _{5})d^{b}(x)]_{j}\}.
\end{eqnarray}
Here and what as follows, the notations are defined in accord with
the previous equation such as $(\partial_\rho\partial_\sigma
+{1\over 2}g_{\rho\sigma} \bar\sqcup)$ as coming from $[(p+q)_\rho
(p+q)_\sigma +{1\over 2}(p+q)^2 g_{\rho\sigma}]$. Similarly, the
renormalized operators $[u(x)u(x)]_{R}$ and $[d(x)d(x)]_{R}$ are
given by

\begin{eqnarray}
\left[ u_{i}^{a}(x)u_{j}^{b}(x)\right] _{R} &=&u_{i}^{a}(x)u_{j}^{b}(x)+%
\frac{G_{F}}{\sqrt{2}}\frac{\mu ^{d-4}}{48\pi ^{2}(d-4)}(\partial
_{\rho
}\partial _{\sigma }+\frac{1}{2}g_{\rho \sigma }\bar\sqcup )\times  \nonumber \\
&&[\gamma ^{\rho }\gamma _{\mu }(A^{u}-B^{u}\gamma
_{5})u^{a}(x)]_{i}[\gamma
^{\sigma }\gamma ^{\mu }(A^{u}-B^{u}\gamma _{5})u^{b}(x)]_{j}.\nonumber\\
&&
\end{eqnarray}

\begin{eqnarray}
\left[ d_{i}^{a}(x)d_{j}^{b}(x)\right] _{R} &=&d_{i}^{a}(x)d_{j}^{b}(x)+%
\frac{G_{F}}{\sqrt{2}}\frac{\mu ^{d-4}}{48\pi ^{2}(d-4)}(\partial
_{\rho
}\partial _{\sigma }+\frac{1}{2}g_{\rho \sigma }\bar\sqcup )\times  \nonumber \\
&&[\gamma ^{\rho }\gamma _{\mu }(A^{d}-B^{d}\gamma
_{5})d^{a}(x)]_{i}[\gamma
^{\sigma }\gamma ^{\mu }(A^{d}-B^{d}\gamma _{5})d^{b}(x)]_{j}.\nonumber\\
&&
\end{eqnarray}

The renormalized composite operators $[\eta _{n}]_{R}$ and $[\eta
_{p}]_{R}$ may now be deduced directly from the above three
equations. This corresponds to introduction of the counter terms
which cancel the subdivergences. We find that $[\eta ]_{R}$ is given
by

\begin{eqnarray}
\lbrack \eta _{p}]_{R} &=&\epsilon ^{abc}[u^{aT}(x)C\gamma _{\mu
}u^{b}(x)]\gamma _{5}\gamma _{\mu
}d^{c}(x)+\frac{G_{F}}{\sqrt{2}}\frac{\mu
^{d-4}\epsilon ^{abc}}{24\pi ^{2}(d-4)}\times  \nonumber \\
&&(\partial _{\rho }^{y}\partial _{\sigma }^{y}+\frac{1}{2}g_{\rho
\sigma }(\bar\sqcup)_{y})\times \{-[u^{aT}(y)C\gamma ^{\sigma
}\gamma _{\mu }\gamma ^{\rho
}u^{b}(y)]\gamma _{5}\gamma _{\mu }d^{c}(x)  \nonumber \\
&&+[u^{aT}(x)C\gamma _{\mu }\gamma ^{\rho }\gamma _{\nu
}(A^{u}-B^{u}\gamma _{5})u^{b}(y)]\gamma _{5}\gamma _{\mu }\gamma
^{\sigma }\gamma ^{\nu
}(A^{d}-B^{d}\gamma _{5})d^{c}(y)  \nonumber \\
&&+[u^{aT}(x)C\gamma _{\mu }\gamma ^{\rho }\gamma _{\nu }(1-\gamma
_{5})u^{b}(y)]\gamma _{5}\gamma _{\mu }\gamma ^{\sigma }\gamma ^{\nu
}(1-\gamma _{5})d^{c}(y)\}|_{y=x}.\nonumber\\
&&
\end{eqnarray}
An analogous expression for $[\eta _n]_{R}$ may be obtained by
exchanging all the $u$'s and $d$'s. We note that, once the
subdivergences are removed in this way, the overall divergence
(associated with the polarization function) becomes local and thus
is removed upon Borel transform (as employed in the context of QCD
sum rules). The difference our result from what we may obtain by
working throughout with the renormalizable $R_\xi$ gauge also
disappears upon Borel transformation.

\subsection{QCD sum rules for the parity-violating couplings}

The various diagrams which we need to consider are illustrated in
Figs. 1, where the propagator with a thick external-line mark is to
be understood as the quark propagator in the presence of an external
vector field.\cite{wen96a,Ioffe} At the quark level, we calculate
the polarization function by considering only the leading two terms
in the operator product expansion $-$ the first
one is associated with the operator $<1>$ and the second with the operator $<%
\bar{q}\sigma _{\mu \nu }q>$. As explained immediately below, the
resultant expressions are already extremely complicated
algebraically, preventing us from carrying out a better and more
complete calculation (unless a considerable amount of time is
invested).

\begin{figure}[h!]
\centering
\includegraphics[width=4in]{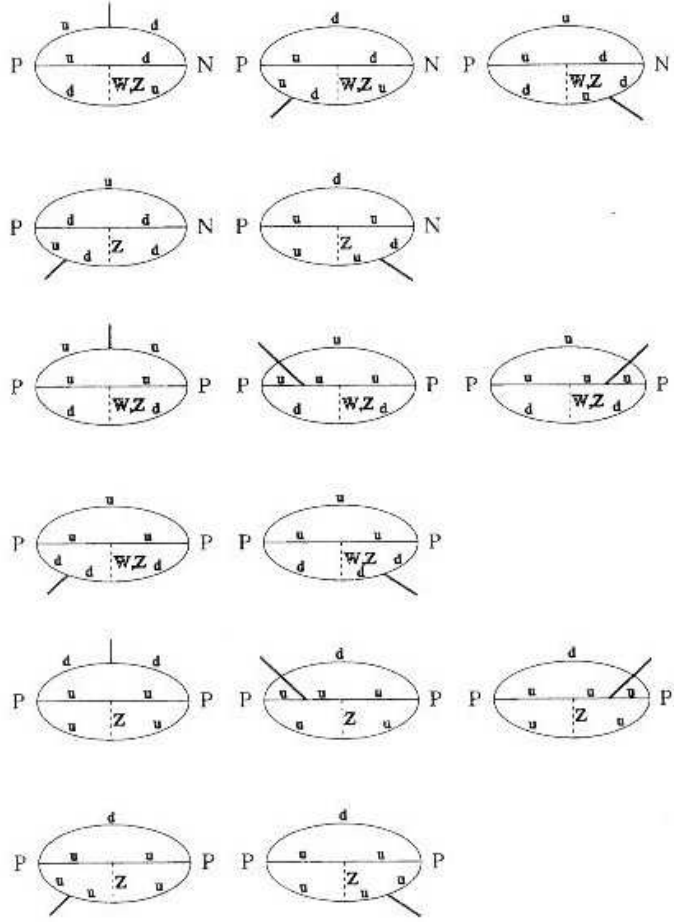}
\caption{Loop diagrams for PV-coupling sum-rule calculations.}
\end{figure}

To get a better feeling towards the extensiveness of the problem, we
may look at the following expression which is associated with the
coefficient of the operator $<1>$:

\begin{eqnarray}
&&\int \frac{d^{d}p_{1}}{(2\pi )^{d}}\frac{d^{d}p_{2}}{(2\pi )^{d}}\frac{%
d^{d}p_{3}}{(2\pi )^{d}}\frac{
p_{1}^{a}p_{2}^{b}(q-p_{3})^{c}(p_{3}-p_{1})^{d}(p_{3}-p_{2})^{e}}{%
p_{1}^{2}p_{2}^{2}(q-p_{3})^{4}(p_{1}-p_{3})^{2}(p_{2}-p_{3})^{2}}\times
\nonumber \\
&& \gamma ^{\mu }\gamma _{a}\gamma _{\alpha }(A^{d}-B^{d}\gamma
_{5})\gamma _{b}\gamma _{\nu }(\gamma _{c}\sigma _{\rho \sigma
}+\sigma _{\rho \sigma }\gamma _{c})\gamma _{\mu }\gamma _{d}\gamma
_{\alpha
}(A^{u}-B^{u}\gamma _{5})\gamma _{e}\gamma ^{\nu },\nonumber\\
&&
\end{eqnarray}
which involves taking the trace of the product of 13 gamma matrices,
a task that needs a good algebraic software package (such as
Mathematica). It also involves three integration variables
$p_{1}$,$p_{2}$ and $p_{3}$ and dimensional regularization is
required during the integration. To work out the problem, we have
chosen to simplify these three-loop expressions by devising
algebraic programs making use of Mathematica. It still takes up
considerable amount of computer time in solving the problem. In
practice, we perform the calculation in two steps: First, we do the
integrations in the order of $p_{1}$ and $p_{2}$, then $p_{3}$. This
needs a liitle program written in Mathematica to handle it (using
the various formulae suitable in $d-$dimensions). The result
contains about a hundred terms with different tensor structures.
Second, we contract the result with the gamma matrices and
simplifying it. For this step we use the Mathematica package
''FeynCalc 1.0'' (written by Rolf Mertig). The final result is

\begin{eqnarray}
&&-\frac{q^{4}}{\pi ^{6}}(-\hat{q}g_{\rho \sigma }+\hat{q}\gamma
_{\rho }\gamma _{\sigma }-\gamma _{\sigma }q_{\rho }+\gamma _{\rho
}q_{\sigma
})[(A^{d}A^{u}+B^{d}B^{u})  \nonumber \\
&&\qquad -(A^{u}B^{d}+A^{d}B^{u})\gamma _{5}]\times \{\frac{1}{384(d-4)^{2}}+%
\frac{1}{27648(d-4)}  \nonumber \\
&&\qquad [-275+108X]+\frac{1}{165888}[4345-27\pi
^{2}-2475X+486X^{2}]\},
\nonumber\\
&&
\end{eqnarray}
where $X=\gamma -\log (4\pi )+\log (-q^{2})$. The presence of the
pole term $X/(d-4)$ indicates that this diagram contains
subdivergences. The subdivergences come from the $p_{1}$ and $p_{2}$
integrals. The removal of these subdivergences is done by making use
of the renormalized operators $[\eta _{p}]_{R}$ and $[\eta
_{n}]_{R}$.

We have examined the question of how to define $\gamma_5$ in
$d-$dimensions, since the definition of $\gamma _{5}$ is a tricky
issue in the $d-$dimension ('t Hooft \& Veltman in 1972). In the
present case, however, what we may do is to ignore this fact and
anti-commute it with all the $\gamma_{\mu }$'s to reach the utmost
right position. This is what we have done in the above expresion,
i.e., by moving all the $\gamma _{5}$ matrix to the right of all
gamma matrices.  After removing all the subdivergences, the
difference between using the proper procedure and using the naive
method turns out to be only a polynomial of $q^{2}$. The difference
does not contain nonlocal terms like $\ln (-q^{2})$ or $1/q^{2}$
which we must handle with care.

At the hadron level, there are four coupling constants for the $\rho
$-meson and two for the $\omega $-meson so that the calculation
should be carried out for the various channels including $\rho
^{+}np$, $\rho ^{-}pn$, $\rho ^{0}pp$, $\rho ^{0}nn$, $\omega pp$,
and $\omega nn$ and suitable linear combinations allow for the
determination of the various PV couplings.

We choose to focus on the antisymmetric part of the sum rules
(proportional to $Z^A_{\mu\nu}$ with $Z^A_{\nu\mu}= -Z^A_{\mu\nu}$),
which already contain enough information to determine the
meson-nucleon couplings. Different tensor structures leads to
different sum rules. After suitably adding and subtracting between
these $\rho NN$ $(\omega NN)$ sum rules we obtain the following
results (in the MS scheme).

\begin{eqnarray}
\frac{\lambda _{N}^{2}h_{\rho }^{0}}{(q^{2}-m^{2})^{2}} &=&-\frac{G_{F}h}{%
\sqrt{2}}\frac{q^{4}}{256\pi ^{6}}\times  \nonumber \\
&&(31\ln (-\frac{q^{2}}{\bar{\mu}^{2}})-6\ln ^{2}(-\frac{q^{2}}{\bar{\mu}^{2}%
}))(\frac{1}{2}+\sin ^{2}\theta _{W}) \\
\frac{\lambda _{N}^{2}h_{\rho }^{1}}{(q^{2}-m^{2})^{2}} &=&\frac{G_{F}h}{%
\sqrt{2}}\frac{q^{4}}{256\pi ^{6}}\times  \nonumber \\
&&\frac{1}{36}(59\ln (-\frac{q^{2}}{\bar{\mu}^{2}})-6\ln ^{2}(-\frac{q^{2}}{%
\bar{\mu}^{2}}))(\frac{1}{3}\sin ^{2}\theta _{W}) \\
h_{\rho }^{2} &=&0 \\
\frac{\lambda _{N}^{2}h_{\rho }^{1^{\prime }}}{2m(q^{2}-m^{2})^{2}} &=&-%
\frac{G_{F}h}{\sqrt{2}}\chi \left\langle \overline{q}q\right\rangle \frac{%
q^{2}}{432\pi ^{4}}\times  \nonumber \\
&&(13\ln (-\frac{q^{2}}{\bar{\mu}^{2}})-3\ln ^{2}(-\frac{q^{2}}{\bar{\mu}^{2}%
}))(\frac{1}{3}\sin ^{2}\theta _{W}) \\
\frac{\lambda _{N}^{2}h_{\omega }^{0}}{(q^{2}-m^{2})^{2}} &=&-\frac{G_{F}h}{%
\sqrt{2}}\frac{q^{4}}{256\pi ^{6}}\times  \nonumber \\
&&(31\ln (-\frac{q^{2}}{\bar{\mu}^{2}})-6\ln ^{2}(-\frac{q^{2}}{\bar{\mu}^{2}%
}))(\frac{1}{2}+\sin ^{2}\theta _{W}) \\
\frac{\lambda _{N}^{2}h_{\omega }^{1}}{(q^{2}-m^{2})^{2}} &=&-\frac{G_{F}h}{%
\sqrt{2}}\frac{q^{4}}{256\pi ^{6}}\times  \nonumber \\
&&\frac{5}{36}(35\ln (-\frac{q^{2}}{\bar{\mu}^{2}})-6\ln ^{2}(-\frac{q^{2}}{%
\bar{\mu}^{2}}))(\frac{1}{3}\sin ^{2}\theta _{W})
\end{eqnarray}
where $\ln (1/\overline{\mu }^{2})=\gamma +\ln (1/4\pi \mu ^{2})$.
The entity $\mu $ is the renormalization scale used in dimensional
regularization. $sin^{2}\theta _{W}$ is the electroweak mixing
parameter in the GSW electroweak theory.

Performing Borel transformation on both sides, taking into account
anomalous dimensions for the various terms in the OPE, and making
use of the continuum approximation for contributions from higher
exited states, we find

\begin{eqnarray}
\frac{\lambda _{N}^{2}h_{\rho }^{0}}{M^{2}}e^{-\frac{m^{2}}{M^{2}}} &=&-%
\frac{G_{F}h}{\sqrt{2}}\frac{M^{6}}{256\pi
^{6}}L^{-\frac{4}{9}}\times
\nonumber \\
&&[-62E_{2}(\frac{W}{M})+24F_{2}(\frac{W}{M},\frac{M}{\bar{\mu}})](\frac{1}{2%
}+\sin ^{2}\theta _{W}) \\
\frac{\lambda _{N}^{2}h_{\rho }^{1}}{M^{2}}e^{-\frac{m^{2}}{M^{2}}} &=&\frac{%
G_{F}h}{\sqrt{2}}\frac{M^{6}}{256\pi ^{6}}L^{-\frac{4}{9}}\times
\nonumber
\\
&&\frac{1}{36}[-118E_{2}(\frac{W}{M})+24F_{2}(\frac{W}{M},\frac{M}{\bar{\mu}}%
)](\frac{1}{3}\sin ^{2}\theta _{W}) \\
h_{\rho }^{2} &=&0 \\
\frac{h_{\rho }^{1^{\prime }}}{2m}\frac{\lambda _{N}^{2}}{M^{2}}e^{-\frac{%
m^{2}}{M^{2}}} &=&-\frac{G_{F}h}{\sqrt{2}}\chi \left\langle \overline{q}%
q\right\rangle \frac{M^{4}}{432\pi ^{4}}L^{-\frac{16}{27}}\times
\nonumber
\\
&&[-13E_{1}(\frac{W}{M})+6F_{1}(\frac{W}{M},\frac{M}{\bar{\mu}})](\frac{1}{3}%
\sin ^{2}\theta _{W}) \\
\frac{\lambda _{N}^{2}h_{\omega }^{0}}{M^{2}}e^{-\frac{m^{2}}{M^{2}}} &=&-%
\frac{G_{F}h}{\sqrt{2}}\frac{M^{6}}{256\pi
^{6}}L^{-\frac{4}{9}}\times
\nonumber \\
&&[-62E_{2}(\frac{W}{M})+24F_{2}(\frac{W}{M},\frac{M}{\bar{\mu}})](\frac{1}{2%
}+\sin ^{2}\theta _{W}) \\
\frac{\lambda _{N}^{2}h_{\omega }^{1}}{M^{2}}e^{-\frac{m^{2}}{M^{2}}} &=&-%
\frac{G_{F}h}{\sqrt{2}}\frac{M^{6}}{256\pi
^{6}}L^{-\frac{4}{9}}\times
\nonumber \\
&&\frac{5}{36}[-70E_{2}(\frac{W}{M})+24F_{2}(\frac{W}{M},\frac{M}{\bar{\mu}})%
](\frac{1}{3}\sin ^{2}\theta _{W})
\end{eqnarray}
The function $L$ ($\equiv \ln (M/\Lambda _{QCD})/\ln
(\bar{\mu}/\Lambda_{QCD})$ with $\Lambda _{QCD}=100\,MeV$ and
$\bar{\mu}=0.5\,GeV$) is introduced to take care of the anomalous
dimensions. Unlike our analysis of the QCD sum rules for the strong
couplings $g_\rho$ and $g_\omega$ \cite{wen96a}, loop integrations
in the weak parity-violating case yields subdivergences in the form
of $p^{2n}[\ln (-p^{2}/\bar{\mu}^{2})]^{2}$ with $n$ some integer
which, upon Borel transform, yield terms proportional to
$M^{2n-2}\ln (M^{2}/\bar{\mu}^{2})$ in the QCD sum rules. Thus, the
value for the dimensional parameter $\bar{\mu}$ is of some numerical
importance. The standard choice $\bar{\mu}=0.5\,GeV$ is used in this
paper.

Furthermore, the quantity $W$ (with the standard choice of
$W=1.45\,GeV$) is the threshold used in the continuum approximation,
in which the contributions due to the excited states and the
continuum are approximated by what may be obtained at the quark
level (by use of QCD). The continuum approximation introduces, into
the sum rules (20)-(25), the functions $E_{i}$ and $F_{i}$ which are
defined in the Appendix.

We should also mention that the QCD sum rules which we have obtained
for PV couplings are basically those of leading order in QCD (with
the nontrivial vacuum structure). It is a tremendous task to try to
include a sufficient number of terms which involves the condensates
of higher dimension in nature. Nevertheless, it is important to
emphasize that the QCD sum rule approach is a definitive (deductive)
procedure for evaluating the various diagrams based on QCD, contrary
to the qualitative or semi-quantitative nature of the earlier DDH
approach \cite{DDH}. In other words, the present QCD sum approach
can be improved upon, albeit fairly complicated, order in order in
QCD and with increasing dimensions.

\section{Numerical analysis}

The input parameters for our numerical analysis are those commonly
adopted in standard QCD sum rule analyses
\cite{wen96a,SVZ,Ioffe,HHK}.

\begin{tabular}{lll}
$\lambda _{N}^{2}=1.2\times 10^{-3}\,GeV^{6},$ & $a=0.546\,GeV^{3},$
&
$b=0.47\,GeV^{4}$ \\
& $\chi =-6\,GeV^{-2},$ & $m_{0}^{2}=0.8\,GeV^{2}$ \\
$h=4.65,$ & $g_{\rho }=2.79,$ & $g_{\omega }=8.37$
\end{tabular}

\noindent Variations in some of these parameter values may result in
errors of numerical importance. Neglect of higher dimensional terms
brings in uncertainties which are difficult to quantify (unless some
such terms can explicitly be taken into account). Nevertheless, we
may use our experience from analyzing other QCD sum rules, such as
those for the strong $\rho NN$ and $\omega NN$ couplings
\cite{wen96a}, to assess these errors.

Before performing the standard numerical analysis, we wish to first
use the QCD sum rules (20)-(25), together with some data from the
nuclear parity violation experiments, to provide some estimates on
the various parity-violating (PV) couplings. We note that both the
couplings $h_\rho^0$ and $h_\omega^0$ receive contributions from
both $W^\pm$ and $Z^0$ exchanges while other PV couplings are
dictated by $Z^0$ exchange. As a result, the couplings $h_{\rho
}^{0}$ and $h_{\omega }^{0}$ are much larger than the other PV
coupling constants, which are suppressed by a factor
$(1/3)\sin^{2}\theta _{W}=0.07$. This means that, as an
approximation, we may neglect the couplings $h_{\rho }^{1}$,
$h_{\rho }^{2}$, $h_{\rho }^{\prime 1}$ and $h_{\omega }^{1}$.
Another important approximate relation is $h_{\rho}^{0}=h_{\omega
}^{0}$.

There are many parity violation measurements in processes involving
complex nuclei, leading to the determination of the quantity
$X_{N}^{P}$, which characterizes the strength of PV interaction.
This quantity, with an experimental value of about $3\times
10^{-6}$\cite{BJ}, can be expressed in terms of the PV coupling
constants:

\begin{equation}
X_{N}^{P}=5.5f_{\pi }-0.25g_{\rho }h_{\rho }^{1}-0.62g_{\rho
}h_{\rho }^{0}-0.05g_{\rho }h_{\rho }^{^{\prime }1}-0.17g_{\omega
}h_{\omega }^{1}-0.19g_{\omega }h_{\omega }^{0}.
\end{equation}
The PV $\pi NN$ coupling $f_{\pi }$ may be obtained via the QCD
sum rule method and it is about $(3.0\pm 0.5)\times
10^{-7}$\cite{Hwang,HHK}. Thus, we have $h_{\rho }^{0}\approx
h_{\omega}^{0}\approx -4.1\times 10^{-7}$, by assuming that the
other PV couplings $(h_{\rho }^{1},h_{\rho}^{2},h_{\rho
}^{^{\prime }1},h_{\omega }^{1})$ are considerably smaller in
comparison.

Analogously, we may also use the expression of the asymmetry
$A_{pp}$, as observed in polarized proton-proton scattering, to
determine the PV couplings:

\begin{equation}
A_{pp}(15MeV)=0.01 g_{\omega}(h_{\omega }^{0}+h_{\omega
}^{1})+0.03g_{\rho} (h_{\rho }^{0}+h_{\rho }^{1}+\frac{h_{\rho
}^{2}}{\sqrt{6}}),
\end{equation}
with an experimental value of $-(1.7\pm 0.85)\times
10^{-7}$\cite{DDH}. It yields a value of about $-(10\pm 5)\times
10^{-7}$ for $h_{\rho}^{0}$ and $h_{\omega}^{0}$, the magnitude of
which is somewhat larger than the previous value but remains
consistent in light of large errors.

We proceed to analyze the various QCD sum rules as a function of the
Borel mass squared $M^2$. The choice of the range for the Borel mass
is guided by the study of the QCD sum rules for other properties of
the nucleon, such as the nucleon mass, the magnetic moments, the
axial couplings, or the strong and weak $\pi NN$ couplings.
Specifically, it has been found in general that the nucleon sum
rules should give rise to reasonable predictions in the Borel mass
range of $0.9\,GeV^2 \le M^2 \le 1.1\, GeV^2$. Such choice of the
Borel mass squared also enable us to estimate the errors of our QCD
sum rule predictions by inferring from the analysis of the other QCD
sum rules for the nucleon. The error could be as large as about 25
\% in some special cases.

\begin{figure}[h!]
\centering
\includegraphics[width=4in]{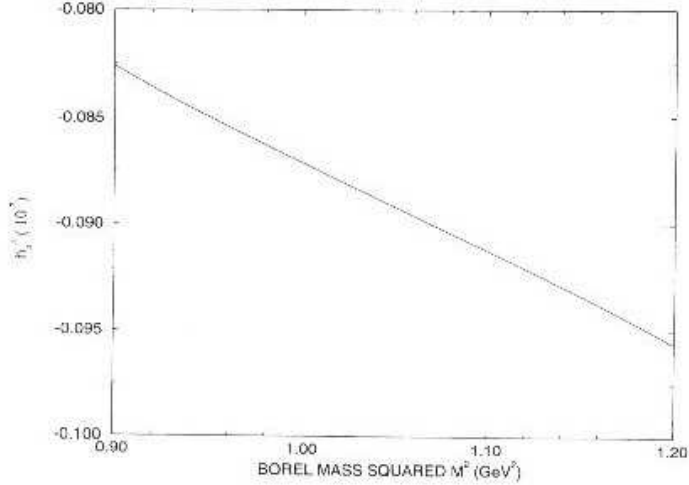}
\caption{The PV coupling $h_\rho^0$ (or $h_\omega^0$) is plotted as
a function of the Borel mass squared $M^2$ in the range of $(0.9 -
1.2)\,GeV^2$.}
\end{figure}

\begin{figure}[h!]
\centering
\includegraphics[width=4in]{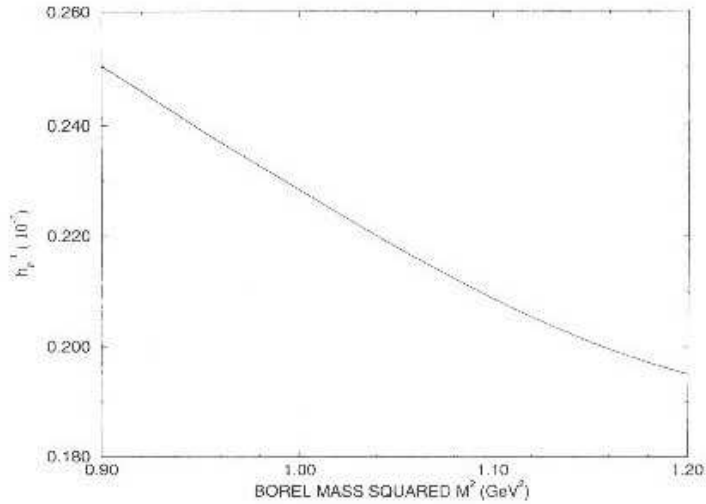}
\caption{The PV coupling $h_\rho^1$ is plotted as a function of the
Borel mass squared $M^2$ in the range of $(0.9 - 1.2)\,GeV^2$.}
\end{figure}

In Figs. 2-5, the various couplings are shown as a function of the
Borel mass squared $M^2$ (in units of $GeV^2$). It is of some
importance to note that the scales in these figures are in fact
different, resulting in errors of different magnitudes. As long as
we have faith in these sum rules when a sufficient number of higher
dimensional terms are included, it makes sense to make
(semi-quantitative) predictions with the aid of only a couple of
leading terms. The following predictions have been obtained in this
way:

\begin{eqnarray*}
h_{\rho }^{0}: &-(6.9\pm 3.6)\times 10^{-7} \\
h_{\rho }^{1}: & -(0.087\pm 0.004)\times 10^{-7} \\
h_{\rho }^{2}: &0 \\
h_{\rho }^{1^{\prime }}: &(0.23\pm 0.02)\times 10^{-7} \\
h_{\omega }^{0}: &-(6.9 \pm 3.6)\times 10^{-7} \\
h_{\omega }^{1}: &-(0.03\pm 0.04)\times 10^{-7}.
\end{eqnarray*}
The large errors, such as those associated with $h_\rho^0$ and
$h_\omega^0$, are caused primarily by the rapid variation of the
prediction in the quoted Borel mass range. The experience with the
QCD sum rule method, such as in the weak parity-violating coupling
$f_{\pi NN}$ \cite{HHK}, suggests that inclusion of the next couple
of terms of higher dimensions could help to smoothen the
rapid-varying behavior of the leading term while yielding
predictions in the same ballpark.

\begin{figure}[h!]
\centering
\includegraphics[width=4in]{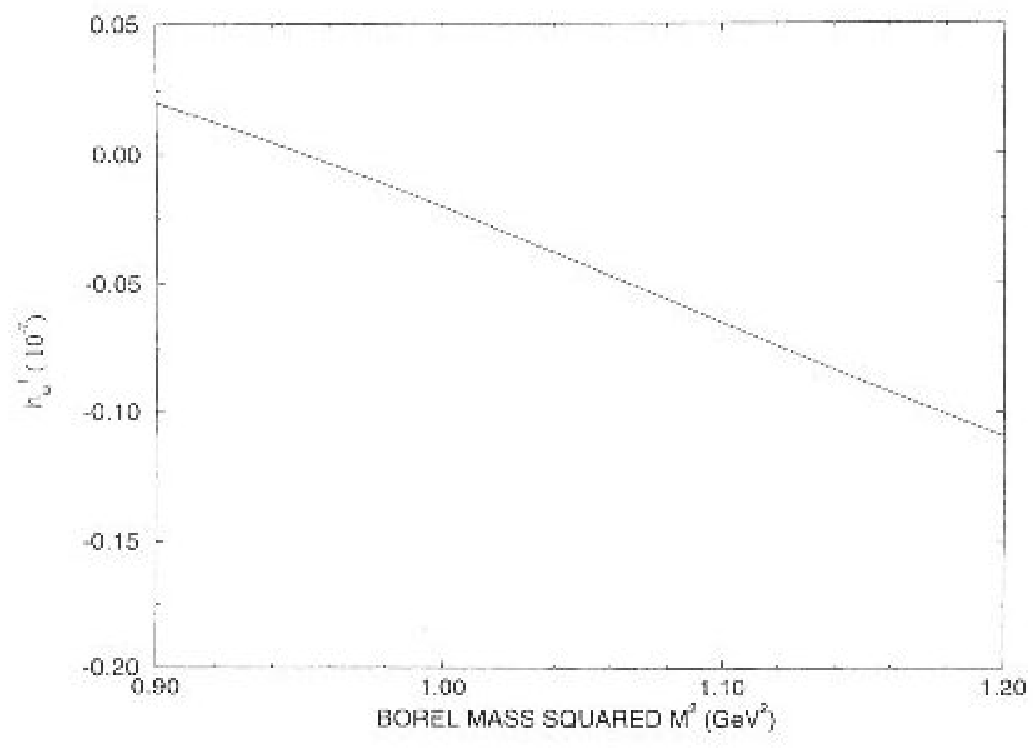}
\caption{The PV coupling $h_\rho^{\prime 1}$ is plotted as a
function of the Borel mass squared $M^2$ in the range of $(0.9 -
1.2)\,GeV^2$.}
\end{figure}

\begin{figure}[h!]
\centering
\includegraphics[width=4in]{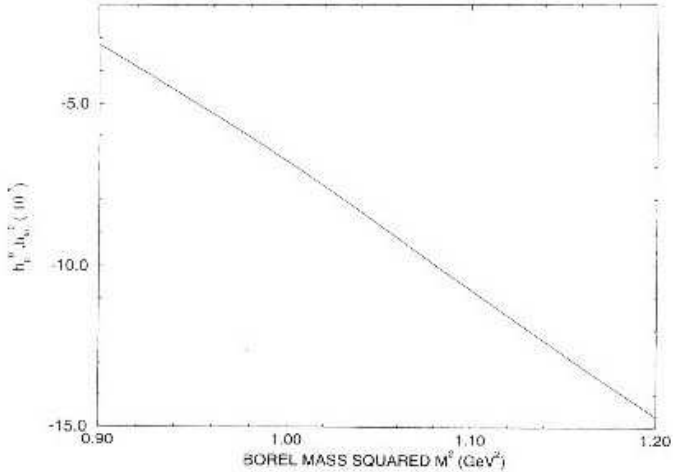}
\caption{The PV coupling $h_\omega^1$ is plotted as a function of
the Borel mass squared $M^2$ in the range of $(0.9 - 1.2)\,GeV^2$.}
\end{figure}

We note that the predicted values for $h_{\rho }^{0}$ and $h_{\omega
}^{0}$ are compatible with the above results as extracted from the
experimental values of $X_N^P$ and $A_{pp}(15\,MeV)$ (together with
the observation from the QCD sum rules that only $h_{\rho }^{0}$ and
$h_{\omega }^{0}$ are dominant).

In Table I, we compare the various predictions on parity-violating
meson-nucleon couplings, including the estimates given by
Desplanques, Donoghue, and Holstein (DDH) \cite{DDH}, the values by
an overall fit to the existing experiments (AH) \cite{AH}, and ours
making use of the QCD sum rules. Our results, if taken seriously,
sharpen the allowed ranges for most of the six couplings. The
overall agreement is good, although we should note that, from the
leading diagrams which we have included, the isotensor coupling
$h_\rho^2$ vanishes (while it fairly sizable in \cite{DDH,AH}) and
$h_\rho^{\prime 1}$ is small but is different from zero (as in
\cite{DDH,AH}). [Both $h_\rho^2$ and $h_\rho^{\prime 1}$ do not
contribute in any major way to the existing nuclear parity violation
observables. See, for example, the expressions for $X_N^P$ and
$A_{pp}(15\, MeV)$.]

\begin{table}[tbp]
\begin{tabular}{|l|l|l|l|}
\hline & DDH (best estimate) & AH (best fit) & QCD sum rules \\
\hline
$f_{\pi }$ & $0\rightarrow 11 (4.6)$ & 0$\rightarrow 11 (2.1)$ & $%
3.0 \pm 0.5 ^*$ \\ \hline
$h_{\rho }^{0}$ & $-31\rightarrow 11 (-11.4)$ & $-31\rightarrow 11 (-5.8)$ & $%
-(6.9 \pm 3.7) $ \\ \hline $h_{\rho }^{1}$ & $-0.4\rightarrow 0
(-0.19)$ & $-0.5\rightarrow 0.4 (-0.22)$ & $- (0.087\pm 0.004)$ \\
\hline $h_{\rho }^{2}$ & $-7.6\rightarrow -11 (-9.5)$ &
$-6.3\rightarrow -10 (-7.1)$ &\qquad $0$ \\ \hline $h_{\rho
}^{\prime 1}$ &\qquad $0$ &\qquad $0$ & $ 0.23\pm 0.02 $ \\ \hline
$h_{\omega }^{0}$ & $-10\rightarrow 5.7 (-1.9)$ & $-12\rightarrow
2.6 (-5.0)$ & $-(6.9\pm 3.7) $ \\ \hline $h_{\omega }^{1}$ &
$-0.8\rightarrow -1.9 (-1.1)$ & $-3.1\rightarrow -1.1 (-2.4)$ & $-
(0.03 \pm 0.04) $ \\ \hline
\end{tabular}
\par
Table 1: PV couplings in units of $10^{-7}$. The numbers in the
brackets are the ``best'' values. $^*$ The prediction is obtained
from \cite{Hwang}.
\end{table}

Owing to the tremendous complications to include a sufficient number
of higher dimensional terms involving the various condensates, we
have relied on the experience of analyzing the other QCD sum rules
for the nucleon in order to assess the uncertainties. While it is
clearly desirable to improve on these derivations (by including more
higher order terms, especially for $h_\rho^0$ and $h_\omega^0$), it
is indeed gratifying to note that our overall predictions are fairly
consistent with earlier results \cite{DDH,AH}.

\section{Summary}

The problem of parity-violating nuclear force has been a very
difficult one. Progresses have been made in the past \cite{DDH,AH}
in the simple meson-exchange picture, in which parity-violating
nuclear force arises from exchange of $\pi $, $\rho $, $\omega $,
or other meson with strong $MNN$ coupling at one vertex and weak
parity-violating $MNN$ coupling at the other vertex. The QCD sum
rule method allows for a determination of the various
parity-violating $MNN$ couplings starting from QCD (with the
nontrivial vacuum) and Glashow-Salam-Weinberg (GSW) electroweak
theory. I believe that the QCD sum rule methods, which make
explicit connection between the underlying theory (i.e. the QCD
and GSW electroweak theory) and the predictions, offer us a
systematic method to tackle. We wish to note that our QCD sum rule
predictions are fairly consistent with earlier
results\cite{DDH,AH,Michael}.

\section*{Acknowledgments}

The authors wish to acknowledge Center for Theoretical Physics of
MIT for the hospitalities extended to them during which the major
part of this project was carried out. This work was supported by
the National Science Council of R.O.C. (through the grant
NSC96-2112-M002-023-MY3 and NSC96-2752-M002-007-PAE).

\section*{Appendix: Borel Transformation}

For the invariant functions $f(p^2)$ appearing in the polarization
function, we may apply the following dispersion relation,

\begin{equation}
f(s)=\frac 1\pi \int_0^\infty \frac{Im\ f(p^{\prime 2})}{p^{\prime \
2}+s}d\ p^{\prime \ 2},\quad s=-p^2,
\end{equation}
with a necessary number of subtractions.

Borel transformation is a general tool employed in the method of QCD
sum rules. It is defined by

\begin{equation}
B\ f(s)=\lim_{
\begin{array}{l}
n,s\rightarrow \infty \\
s/n=M^{2}
\end{array}
}\frac{s^{n+1}}{n!}\ (-\frac{d}{d\ s})^{n}f(s)
\end{equation}

Apply Borel transformation to the dispersion relation we obtain

\begin{equation}
B\ f(s)=\frac 1\pi \int_0^\infty e^{-p^2/M^2}Im\ f(p^2)\ dp^2.
\end{equation}

In general, we may take into account the contribution due to exited
states and assume that it is equal to the quark-level contribution
of $f(p^{2})$ starting from some cutoff value $p^{2}=W^{2}$. That
is, we use a modified Borel transformation $B_{W}$ with the
definition:

\begin{equation}
B_W\ f(s)=\frac 1\pi \int_0^{W^2}e^{-p^2/M^2}Im\ f(p^2)\ dp^2,
\end{equation}
which involves some cutoff $W$.

We list all the modified Borel transformation functions which we
need in the text.

\begin{eqnarray}
-M^{2}E_{0}(\frac{W}{M}) &\equiv & B_{W}\ \ln \frac{s}{\mu ^{2}}
\nonumber
\\
&=&-M^{2}(1-e^{-W^{2}/M^{2}}) \\
M^{4}E_{1}(\frac{W}{M}) &\equiv & B_{W}\ s\ln \frac{s}{\mu ^{2}}
\nonumber
\\
&=&M^{4}[1-e^{-W^{2}/M^{2}}(1+\frac{W^{2}}{M^{2}})] \\
-2M^{6}E_{2}(\frac{W}{M}) &\equiv & B_{W}\ s^{2}\ln \frac{s}{\mu
^{2}}
\nonumber \\
&=&-2M^{6}[1-e^{-W^{2}/M^{2}}(1+\frac{W^{2}}{M^{2}}+\frac{W^{4}}{2M^{4}})] \\
2M^{4}F_{1}(\frac{W}{M},\frac{M}{\mu }) &\equiv & B_{W}\ s\ln ^{2}\frac{s}{%
\mu ^{2}}  \nonumber \\
&=&2M^{4}\{1-\gamma -e^{-W^{2}/M^{2}}+Ei(-\frac{W^{2}}{M^{2}})  \nonumber \\
&&+[1-e^{-W^{2}/M^{2}}(1+\frac{W^{2}}{M^{2}})]\ln \frac{M^{2}}{\mu ^{2}}\} \\
-4M^{6}F_{2}(\frac{W}{M},\frac{M}{\mu }) &\equiv & B_{W}\ s^{2}\ln ^{2}\frac{%
s}{\mu ^{2}}  \nonumber \\
&=&-4M^{6}\{\frac{3}{2}-\gamma -\frac{3}{2}e^{-W^{2}/M^{2}}(1+\frac{W^{2}}{%
3M^{2}})+Ei(-\frac{W^{2}}{M^{2}})  \nonumber \\
&&+[1-e^{-W^{2}/M^{2}}(1+\frac{W^{2}}{M^{2}}+\frac{W^{4}}{2M^{4}})]\ln \frac{%
M^{2}}{\mu ^{2}}\}
\end{eqnarray}
Here we have adopted the conventional symbols $E_{0}$,$E_{1}$ and
$E_{2}$, and introduced two new symbols $F_{1}$ and $F_{2}$. The
exponential integral function $Ei(x)$ is defined by

\begin{equation}
Ei(x)=-\int_{-x}^{\infty }\frac{e^{-t}}{t}dt
\end{equation}

\end{document}